\documentclass[doublecol, a4paper]{epl2}

\usepackage{amsmath}
\usepackage{graphicx}
\usepackage{bm}
\usepackage{gensymb} 
\bibstyle{eplbib} 
\usepackage[hidelinks]{hyperref}

\renewcommand{\revision}[1]{{\color{black}}{#1}}

\title{Synchronization of particle motion in compressed two-dimensional plasma crystals}

\author{I.~Laut\inst{1}\thanks{E-mail: \email{ingo.laut@dlr.de}} \and 
        C.~R\"{a}th\inst{1} \and 
        S.~Zhdanov\inst{2} \and 
        V.~Nosenko\inst{1} \and 
        L.~Cou\"edel\inst{3} \and 
        H.~M.~Thomas\inst{1}}
\shortauthor{I. Laut \etal}

\institute{                    
  \inst{1} Deutsches Zentrum f\"{u}r Luft- und Raumfahrt, Forschungsgruppe Komplexe Plasmen, 82234 We{\ss}ling, Germany \\
  \inst{2} Max Planck Institute for Extraterrestrial Physics, 85741 Garching, Germany \\
  \inst{3} CNRS, Aix-Marseille Universit\'e, Laboratoire de Physique des Interactions Ioniques et Mol\'eculaires, 13397 Marseille, France
}


\abstract
{
The collective motion of dust particles during the mode-coupling induced melting of a two-dimensional plasma crystal is explored in molecular dynamics simulations. The crystal is compressed horizontally by an anisotropic confinement. This compression leads to an asymmetric triggering of the mode-coupling instability which is accompanied by alternating chains of in-phase and anti-phase oscillating particles. 
A new order parameter is proposed to quantify the synchronization with respect to different directions of the crystal. Depending on the orientation of the confinement anisotropy, mode-coupling instability and synchronized motion are observed in one or two directions. Notably, the synchronization is found to be direction-dependent. The good agreement with experiments suggests that the confinement anisotropy can be used to explain the observed synchronization process.
}

\begin{document}

\maketitle

\section{Introduction}
Weakly ionized gases containing micron-sized dust particles are called complex (dusty) plasmas. In the plasma the particles charge up and self-arrange enabling formation of strongly coupled and highly ordered quasicrystalline phases \cite{ikezi1986, thomas1994, hayashi1994, chu1994, thomas1996} (analogous to colloids \cite{morfill2009}), called \emph{plasma crystals} \cite{thomas1994, samsonov2001, nunomura2000}. In ground-based experiments these crystals are typically composed of plastic microspheres that are injected into a plasma created by a radio frequency discharge. The particles charge up negatively and levitate in the plasma sheath region above the lower electrode where they can form a horizontal two-dimensional (2D) monolayer under adequate experimental conditions \cite{thomas1994, chu1994}.  Many dynamical processes can be studied rigorously in plasma crystals, in particular, linear \cite{fortov2000, misawa2001} and nonlinear waves \cite{samsonov2002}, resonance effects \cite{williams2014}, dynamics of dislocations \cite{nosenko2007, nosenko2008dislocation, zhdanov2011spontaneous} and crystal plasticity \cite{durniak2011, durniak2013}. 

As in many physical, astrophysical and biological systems \cite{zwicky1933}, cooperative particle motion is an exceptionally important element of self-organization in complex plasmas. In particular, synchronized motion of particle chains was recently discovered in plasma crystals \cite{couedel2014}. Synchronization processes in large systems of oscillators have been studied in chemistry, physics and engineering \cite{kuramoto1984chemical}, 
and the behavior of chirping crickets \cite{walker1969}, 
or superconducting Josephson junctions \cite{wiesenfeld1996} can be described by the Kuramoto model of globally coupled oscillators \cite{kuramoto1984cooperative} which can be solved analytically in a mean-field approach.

In a plasma crystal, the particle-particle interaction is strongly influenced by the surrounding plasma. While the interaction in the bulk plasma is well described by a Yukawa potential \cite{ikezi1986}, the strong ion flow in the plasma sheath region distorts the screening cloud \cite{schella2013, laut2014}. This plasma \emph{wake} below the particles adds an attractive component to the interaction \cite{melzer1999} which was described theoretically as a pointlike positive effective charge below each particle \cite{ivlev2000}. Due to the finite vertical confinement of a 2D plasma crystal, there is an out-of-plane wave mode which has an optical dispersion relation in addition to the two in-plane modes with acoustic dispersion. If the vertical confinement is smaller than a critical value, the longitudinal in-plane mode and the out-of-plane mode intersect and form an unstable hybrid mode in the vicinity of the intersection. During this \emph{mode-coupling instability} (MCI), energy is continuously transferred from the flowing ions to the crystal, breaking the crystalline order if the damping rate is small enough \cite{zhdanov2009, couedel2011}. 

Due to the lattice symmetry, MCI in the shallow crossing regime is equally strong in three directions in a perfect hexagonal lattice \cite{couedel2011}. In the experiment of ref.~\cite{couedel2014}, however, the instability was dominant in one direction. Synchronized motion of particle chains was observed. The process of synchronization was measured by calculating the Shannon entropy of the instantaneous phases of neighboring particles as well as the distribution of frequencies. In ref.~\cite{couedel2014}, an inhomogeneity of the horizontal confinement was suggested to be a reason for this asymmetry in the crystal, but it was not possible to study the origin of the deformation of the crystal in detail. 

The influence of an anisotropy in the horizontal confinement on a rotating plasma crystal was studied in ref.\cite{schablinski2014}. It was shown that even small anisotropies may considerably affect the dynamical behavior of the system.

In this paper, we demonstrate with simulations that an anisotropy of the horizontal confinement can cause an asymmetric triggering of MCI. At the onset of the instability, synchronized particle motion is characterized by a new order parameter that is sensitive to the direction of the synchronization pattern. Depending on the orientation of the confinement anisotropy, MCI and synchronized motion are observed in one or two directions.

\section{Experiment}
\begin{figure}
\includegraphics[width=\columnwidth]{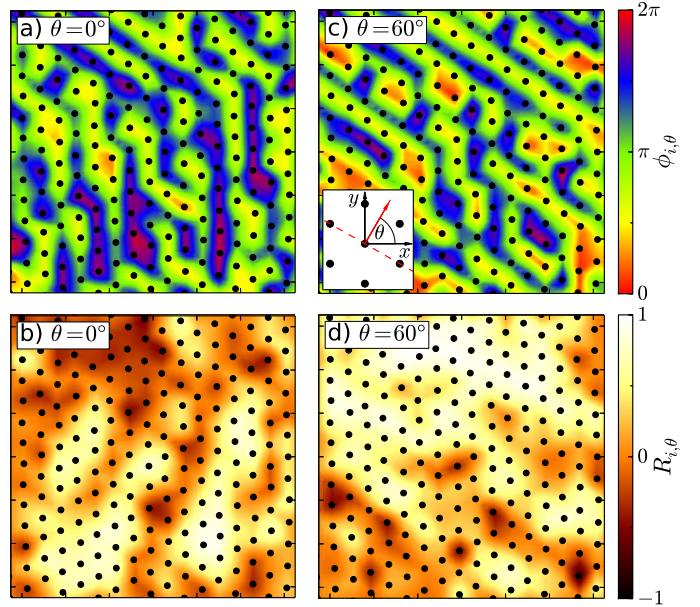}
\caption{
(a) Map of the instantaneous phases $\phi_{i, \theta=0\degree}$ of the particle oscillations at time $t=2.5\un{s}$. The particle positions are indicated by black dots. \revision{The phases of the particles are interpolated between the particle positions in order to be visualized in a map. Lines of particles with similar phases appear as stripes.} The window has a side length $6.5\un{mm}$. \revision{(b) Map of the order parameter $R_{i, \theta=0\degree}$ [see eq.~(\ref{eq:OrderParameter})] at the same time step.}
(c), (d) The same for $\theta=60\degree$. The inset shows the reference frame. The direction denoted by angle $\theta$ is shown as a red arrow, the line perpendicular to it as a dashed line.
} \label{fig:1}
\end{figure}

The experiment of ref.~\cite{couedel2014} will be briefly outlined below. Argon plasma was produced using a capacitively coupled radio frequency discharge at 13.56\,MHz with a forward power of 12\,W. The microparticles formed a monolayer with mean interparticle distance $a = (480 \pm 10)\un{\mu m}$. 
The particle $x$ and $y$ positions were obtained with subpixel accuracy from a top-view camera operating at $250$ frames per second. The axes were chosen as depicted in the inset of fig.~\ref{fig:1}. The gas pressure was reduced from $0.94\un{Pa}$ to $0.92\un{Pa}$ to initiate the MCI. 

The spectral distribution of particle velocity fluctuations [see eq.~(\ref{eq:VelocityFluctuations})] in $\vect{k}$-space is highly anisotropic \cite{couedel2014}. As can be seen in fig.~\ref{fig:2}(c), bright 'hot spots', the fingerprints of the developed MCI, appear in two directions, in contrast to a perfect hexagonal crystal where the three directions are equally strong \cite{couedel2011}. 

\section{Simulation}
Molecular dynamics simulations have proven to be an adequate tool to study and compare a wide range of experimental conditions \cite{totsuji2001, ivlev2003, rocker2014nonlinear, ott2014}. The equations of motion read
\begin{equation} \label{eq:equation_of_motion}
m\ddot{\vect{r}}_i + m \nu \dot{\vect{r}}_i 
 = \sum_{j \neq i} \vect{F}_{ji} 
 + \vect{C}_i 
 + \vect{L}_i \,,
\end{equation}
where $\vect{r}_i$ is the position of particle $i$,
$m$ the particle mass and $\nu $ the damping rate.

The force exerted by particle $j$ (and its wake) on particle~$i$ is
\begin{equation}
 \begin{split}
 \vect{F}_{ji}
  = & \frac{Q^2}{r^2_{ji}} \exp \left( -\frac{r_{ji}}{\lambda} \right) 
      \left( 1 + \frac{r_{ji}}{\lambda} \right)
      \frac{\vect{r}_{ji}}{r_{ji}} \\
    & -\frac{q|Q|}{r^2_{w_{ji}}} \exp \left( -\frac{r_{w_{ji}}}{\lambda} \right) 
      \left( 1 + \frac{r_{w_{ji}}}{\lambda} \right) 
      \frac{\vect{r}_{w_{ji}}}{r_{w_{ji}}} \,,
 \end{split}
\end{equation}
where $Q<0$ is the particle charge, $\lambda$ is the screening length, 
$\vect{r}_{ji} = \vect{r}_{i} - \vect{r}_{j}$ and 
$\vect{r}_{w_{ji}} = \vect{r}_{i} - (\vect{r}_{j} - \delta \vect{e}_z)$. \revision{To model the ion wake effect a positive 'extra charge' $q$ ($0 < q < |Q|$) is added a fixed distance $\delta$ ($\delta < \lambda$) below each particle. Note that since in general $\vect{r}_{w_{ji}} \neq \vect{r}_{w_{ij}}$, the forces are nonreciprocal due to the ion wake effect. The ion wake (described in detail in \cite{morfill2009}) is known to be responsible for triggering the MCI \cite{couedel2011}. } 

\revision{To form a monolayer, the equally charged particles have to be confined vertically as well as horizontally. In the experiment, the confinement can be controlled, e.g., by varying the discharge power or gas pressure \cite{couedel2011}. In simulations it is treated as a tunable parameter, allowing us to control the crystal stability and anisotropy effects.} The anisotropic parabolic confinement force in the horizontal plane is characterized by confinement parameter $\Omega_\parallel = 2\pi f_\parallel$ acting in the direction of angle $\alpha$ \revision{(measured from the $x$-axis)}, and $\Omega_\perp = 2\pi f_\perp$ that is perpendicular to it. Thus, 
\begin{equation} \label{eq:HorizontalConfinement}
\vect{C}_i = - 
 \begin{pmatrix}
  \Omega_s^2 x_i  + \Omega_a^2 ( x_i \cos 2\alpha + y_i \sin 2\alpha ) \\
  \Omega_s^2 y_i  + \Omega_a^2 ( x_i \sin 2\alpha - y_i \cos 2\alpha ) \\
  \Omega_z^2 z_i 
 \end{pmatrix} ,
\end{equation}
where $\Omega_s$ and $\Omega_a$ are the symmetric and asymmetric contributions to the horizontal confinement, and $\Omega_z$ the vertical confinement parameter. The symmetric and asymmetric contributions can be expressed as $\Omega_{s, a}^2 = (\Omega_\parallel^2 \pm \Omega_\perp^2)/2$. The orientation of the confinement anisotropy can thus be changed without changing the choice of the axes depicted in the inset of fig.~\ref{fig:1}, \revision{leading to a horizontal compression of the crystal in the direction denoted by the angle $\alpha$.}

The particles are also coupled to a Langevin heat bath of temperature $T = 300\un{K}$,
\begin{equation}
\langle \vect{L}_i(t) \rangle = 0 \,, \quad
\langle \vect{L}_i(t + \tau) \vect{L}_j(t) \rangle 
 = 2 \nu m T \delta_{ij} \delta(\tau) \,.
\end{equation}
$\delta_{ij}$ is the Kronecker delta and $\delta(\tau)$ is the delta function.

In a simulation run, a system of 16384 particles, each with a mass $m = 6.1 \times 10^{-13}\un{kg}$ and charge $Q=-19000e$, is first equilibrated at $f_\parallel = f_\perp = 0.145\un{Hz}$ and a large vertical confinement $f_z = \Omega_z/2\pi = 23\un{Hz}$ that prevents the onset of MCI. When a crystal is formed in the center of the monolayer, the horizontal frequencies are changed to $f_\parallel = 0.156\un{Hz}$ and $f_\perp = 0.137\un{Hz}$ to introduce an anisotropy. After equilibration, the vertical confinement is finally reduced to $f_z = 20\un{Hz}$ to trigger the instability, this moment corresponds to $t=0$. Because of the sixfold symmetry of the crystal, it is sufficient to study the orientation of the confinement anisotropy in the range $0\degree \le \alpha \le 30\degree$. Here, two simulations with $\alpha = 30\degree$ and $\alpha = 0\degree$ are considered. The damping rate is assumed to be $\nu = 1.26\un{s^{-1}}$, the screening length is $\lambda=380\un{\mu m}$. A pointlike wake charge $q = 0.2|Q|$ is a distance $\delta = 0.3\lambda$ below each particle.

\section{Analysis methods}
The radial pair correlation function in the horizontal plane, $g(\vect{r})$, is used to measure the inhomogeneity in the hexagonal lattice. An ellipse is fitted to the first six peaks of $g(\vect{r})$. The tilt angle $\beta$ and the eccentricity $\epsilon$ are used to quantify the deformation of the crystal. 

The particle current \cite{donko2008} for the longitudinal in-plane mode is defined as
\begin{equation}
\label{eq:VelocityFluctuations}
V(\vect{k}, t) = \sum_j v_j^{\vect{k}}(t) e^{-i \vect{k} \cdot \vect{r}_j} \,,
\end{equation}
where $v_j^{\vect{k}}(t)$ is the component of the velocity of particle $j$ at time $t$ parallel to wave vector $\vect{k} = (k_x, k_y)$. The particle current fluctuation spectra of the longitudinal mode $V(\vect{k}, f)$ are then calculated using the Fourier transform. To show the spectra in the $xy$ plane, $V(\vect{k}, f)$ is integrated over a frequency range $14\un{Hz}<f<18\un{Hz}$ centered on the hybrid frequency $f_\mathrm{hyb} = (16\pm1)\un{Hz}$. The MCI, where the out-of-plane mode couples to the longitudinal in-plane mode, appears as hot spots in the spectra of both modes \cite{couedel2011}. The out-of-plane mode is not considered since it is not available for the experimental data. In the simulations, the integrated spectrum of the out-of-plane mode is very similar to that of the longitudinal mode. The border of the first Brillouin zone is calculated from the static structure factor $S(\vect{k}) = N^{-1} \langle \sum_{l, m} e^{i \vect{k} \cdot (\vect{r}_l - \vect{r}_m)}\rangle$, where $N$ is the number of particles, the sum runs over all pairs of particles, and the averaging is performed over time. 

\begin{figure}
\includegraphics[width=\columnwidth]{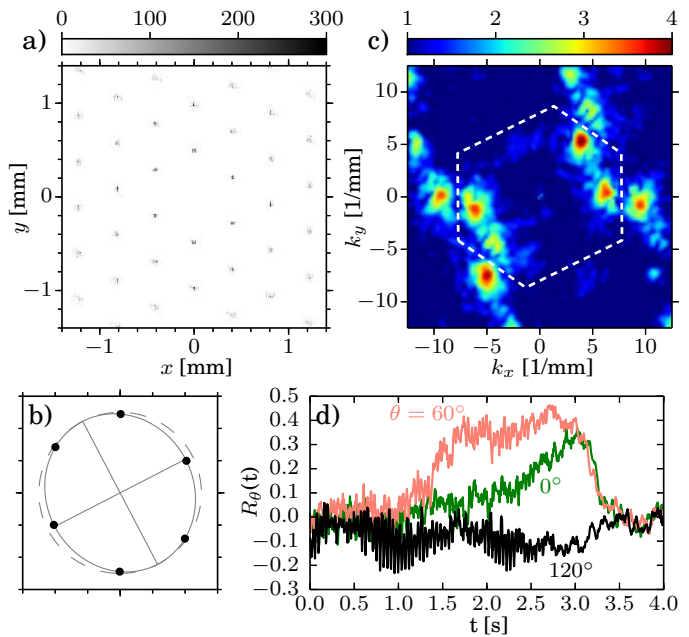}
\caption{
Synchronized particle motion at the onset of mode-coupling induced melting of 2D plasma crystal.
(a)~Pair correlation $g(\vect{r})$ in the horizontal plane at $t=0$. 
(b)~$g(\vect{r})$ in a smaller window of side length $1.2\un{mm}$ with the first peaks shown as solid circles. An ellipse (solid line) is fitted to the positions of the peaks, its semiaxes of length $A=(0.500\pm0.014)\un{mm}$ and $B=(0.454\pm0.011)\un{mm}$ are also shown as solid lines. The tilt angle of the ellipse is $\beta=(27\pm2)\degree$. The dashed circle with a radius identical to $A$ is shown to guide the eye. 
(c)~Integrated particle current fluctuation spectrum of the longitudinal mode in the $k_xk_y$-plane, with arbitrary units in a logarithmic scale, calculated from the first $3.2\un{s}$ of the data. The border of the first Brillouin zone is shown as a dashed line.
(d)~Order parameter $R_\theta(t)$ measuring the degree of synchronization of the particles as a function of time $t$, see eq.~(\ref{eq:OrderParameter}). } 
\label{fig:2}
\end{figure}

The chains of synchronized particle motion (see fig.~\ref{fig:1}) cannot be characterized using the Kuramoto order parameter $r e^{i \psi} = (1/N) \sum_j e^{i \phi_j}$ \cite{kuramoto1984chemical}, because neighboring chains tend to be in antiphase. The contributions to the order parameter would thus cancel even in the presence of a synchronization pattern. Therefore, we define a local order parameter as
\begin{equation}\label{eq:OrderParameter}
R_{i, \theta}(t) = 
 \frac{1}{\mathrm{nn}} 
  \left( \sum_{j =1}^\mathrm{nn} \left[ (-1)^{k_j} 
  \cos(\phi_{j, \theta} - \phi_{i, \theta}) \right] \right) ,
\end{equation}
where $\phi_{i, \theta}$ is the phase of the oscillation of particle $i$ in the direction denoted by angle $\theta$ at time $t$ and nn is the number of nearest neighbors. $k_j = 0$ if particle $j$ is on the line passing through particle $i$ perpendicular to direction denoted by $\theta$, and $k_j = 1$ otherwise\footnote{Since the crystal is highly ordered, the definition of particle lines is straightforward. We consider two neighboring particles $i$ and $j$ to be on a line if the angle between the $i-j$ bond and the line is smaller than $30\degree$.}. In the inset of fig.~\ref{fig:1}, the direction denoted by $\theta=60\degree$ is indicated by an arrow and the line perpendicular to it by a dashed line. The cosine of the phase differences are thus added for nearest neighbors on the same line and subtracted for nearest neighbors on the subsequent lines, leading to $R_{i, \theta} = 1$ if particle $i$ is in a region with perfect alternating in-phase and out-of-phase oscillating lines of particles. In the opposite case, $R_{i, \theta} = -1$. If there is no phase relation, $R_{i, \theta} \simeq 0$.

The instantaneous phase $\phi_{i, \theta}$ is calculated from the projection of position $\vect{r}_i$ in the horizontal plane onto the direction denoted by $\theta$. The instantaneous deviation from the time-averaged particle position is obtained with a sliding window of length $0.2\un{s}$. The phase is then assumed to grow linearly by $2\pi$ between each maximum of the deviation. 
An order parameter for the system is calculated by averaging over all particles $R_\theta(t) = \langle R_{i, \theta}(t) \rangle_i$. The three main directions of the crystal, $\theta = 0\degree$, $60\degree$ and $120\degree$, are considered. 

In ref.~\cite{fukuda2005} a local order parameter was used to increase the resolution for a system where the number of oscillators is small. The local order parameter proposed here is sensitive to the orientation of the synchronization pattern. In fig.~\ref{fig:1}, maps of $\phi_{i, \theta}$ and $R_{i, \theta}$ are shown for the experimental data at a characteristic time $t=2.5\un{s}$ for $\theta = 0\degree$ and $\theta = 60\degree$. Lines with two different orientations are apparent for $\phi_{i, \theta=0\degree}$, see fig.~\ref{fig:1}(a). The corresponding order parameter $R_{i, \theta=0\degree}$ [see fig.\ref{fig:1}(b)] is sensitive to the lines that are oriented along the $y$-axis which are located in the lower part of the inspection window. For $\theta=60\degree$ [figs.~\ref{fig:1}(c) and (d)], the largest values of the order parameter are concentrated in the upper part of the window.

\section{Results}
The experimental data of ref.~\cite{couedel2014} is analyzed in a window containing about 800 particles near the center of the crystal. The pair correlation $g(\vect{r})$ is shown in fig.~\ref{fig:2}(a). A deviation from a perfect hexagonal structure can clearly be seen. In fig.~\ref{fig:2}(b), an ellipse is fitted to the first peaks of $g(\vect{r})$, its tilt angle is $\beta = (27\pm2)\degree$. The value of the eccentricity is $\epsilon = 0.42\pm0.07$. 

The phases $\phi_{i, \theta}$ are calculated for a smaller window of side length $6.5\un{mm}$ containing about 230 particles. In this region synchronized particle motion was observed. As can be seen in fig.~\ref{fig:2}(d), the order parameter $R_\theta$ has significant positive values for $\theta = 0\degree$ and $\theta = 60\degree$. In the latter case, $R_{\theta=60\degree}$ increases between $t\simeq1\un{s}$ and $t\simeq2\un{s}$ and then saturates at a value of $R_{\theta=60\degree} \simeq 0.4$. At $t\simeq3.2\un{s}$ the crystal melts and the order parameter drops back to zero. For $\theta=0\degree$, $R_{\theta=0\degree}$ increases much more slowly in a time interval $1\un{s}<t<3\un{s}$ before also decreasing again when the crystal melts. $R_{\theta=120\degree}$ becomes slightly negative during the phase of synchronized motion in the other directions (see supplementary movie \texttt{mci-synchronization.mp4} for the time evolution of the order parameter).

\begin{figure}
\includegraphics[width=\columnwidth]{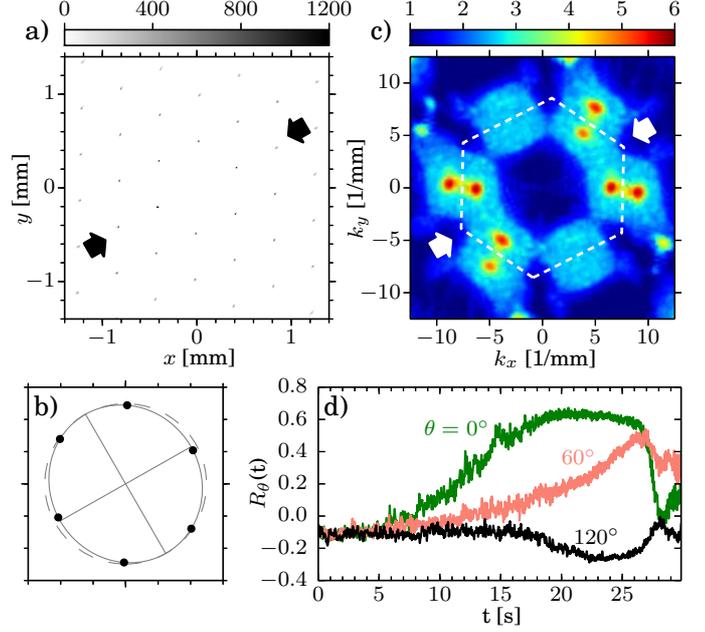}
\caption{
Same as fig.~\ref{fig:2}, but for a molecular dynamics simulation of a crystal with an  anisotropy in the horizontal parabolic confinement. The direction of the largest confinement frequency $f_\parallel = 0.156\un{Hz}$, given by angle $\alpha=30\degree$ \revision{[see eq.~(\ref{eq:HorizontalConfinement})]}, is indicated by arrows in (a) and (c), the frequency in the perpendicular direction is $f_\perp = 0.137\un{Hz}$. The vertical confinement frequency is $f_z = 20\un{Hz}$. In (b), the semiaxes of the ellipse fitted to the first peaks of $g(\vect{r})$ are of length $A=(0.498\pm0.004)\un{mm}$ and $B = (0.464\pm0.003)\un{mm}$. The tilt angle of the ellipse is $\beta = (29.7\pm0.5)\degree$. In (c), the first $25\un{s}$ of the data are used to calculate the spectrum. }
\label{fig:3}
\end{figure}

In the first simulation, the crystal was compressed at an angle of $\alpha = 30\degree$. The region of interest is chosen to be of the same size as in the experiments. The pair correlation $g(\vect{r})$ is shown in fig.~\ref{fig:3}(a). An ellipse is fitted to the first peaks of $g(\vect{r})$ [see fig.~\ref{fig:3}(b)], the value of the tilt angle, $\beta=(29.7\pm0.5)\degree$, is close to the experiment, the value of the eccentricity $\epsilon = 0.36\pm0.03$ is slightly smaller. The integrated spectrum of the longitudinal mode [see fig.~\ref{fig:3}(c)] shows bright hot spots in two main directions of the crystal. The hot spot at $\theta=120\degree$ that would be expected for a perfect hexagonal lattice is almost absent.

The main characteristics of the synchronization process in the experiment are recovered in the simulations, albeit with the roles of $\theta=0\degree$ and $\theta=60\degree$ interchanged: As can be seen in fig.~\ref{fig:3}(d), the order parameter saturates at $R_{\theta=0\degree} \simeq 0.6$ rather quickly in one direction, while it follows more slowly in the other. At $t \simeq 27\un{s}$, both order parameters decrease. 
For $\theta=120\degree$, the order parameter decreases to about $-0.2$ to $-0.3$.
Note that the time scale is larger than in the experiment.

The order parameter is slightly negative even at $t=0$. This could be explained by noting that before the onset of MCI, the particle movement is slightly correlated between nearest neighbors due to their mutual repulsion. This small positive correlation leads to a negative $R_\theta$, since the phase differences to the four neighbors on the next lines are subtracted from the phase differences to only two neighbors on the same line. The decrease of $R_{\theta=120\degree}$ during the period of synchronization can be explained by the fact that the three directions of the projection are not orthogonal. Consequently, alternating lines of in-phase and anti-phase particles in one direction lead to a negative order parameter in the other directions.

\begin{figure}
\includegraphics[width=\columnwidth]{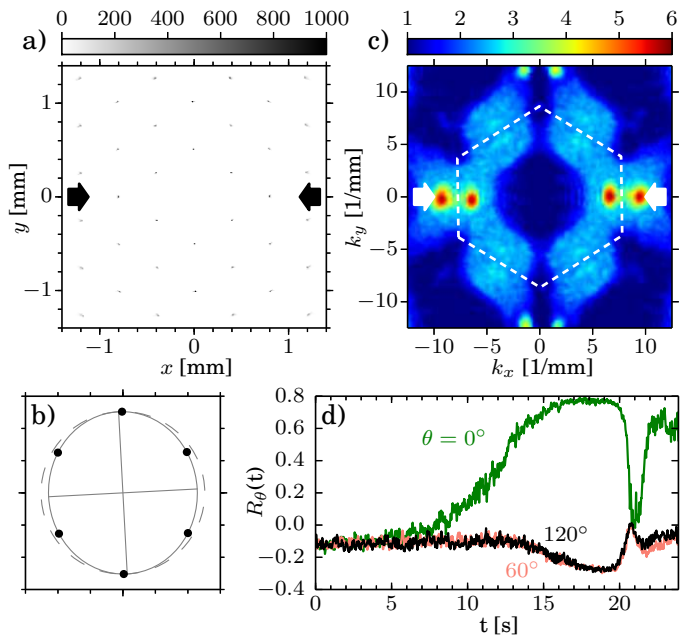}
\caption{
Same as fig.~\ref{fig:3}, but for a different direction of the anisotropy in the horizontal confinement,  $\alpha=0\degree$. In (b), the semiaxes of the ellipse fitted to the first peaks of $g(\vect{r})$ are of length $A=(0.504\pm0.004)\un{mm}$ and $B = (0.458\pm0.004)\un{mm}$. The tilt angle of the ellipse is $\beta = (3.0\pm0.5)\degree$. In (c), the first $20\un{s}$ of the data are used to calculate the spectrum. }
\label{fig:4}
\end{figure}

In a second simulation, the orientation of the confinement anisotropy was $\alpha = 0\degree$. The pair correlation $g(\vect{r})$ and the ellipse fitted to the first six peaks of $g(\vect{r})$ are shown in fig.~\ref{fig:4}(a) and (b). As expected, the ellipse is tilted by only a small angle of $(3.0\pm0.5)\degree$. The eccentricity $\epsilon = 0.42\pm0.03$ is larger than in the first simulation. The integrated particle fluctuation spectrum [fig.~\ref{fig:4}(c)] shows that the MCI is dominant only in the $x$ direction. 

Here, synchronized motion is only observed in $\theta = 0\degree$ direction, see fig.~\ref{fig:4}(d). The corresponding order parameter increases between $t\simeq5\un{s}$ and $t\simeq15\un{s}$ and subsequently saturates at $R_{\theta = 0\degree}\simeq0.8$. $R_\theta$ decreases to negative values (about $-0.2$ to $-0.3$) in the other two directions $\theta = 60\degree$ and $\theta = 120\degree$. 

At $t\simeq20\un{s}$, $R_{\theta=0\degree}$ decreases and increases again at $t\simeq21.5\un{s}$. This can be understood as follows. As observed in \cite{rocker2014nonlinear}, a molecular dynamics simulation of MCI does not lead to a complete melting of the crystal but rather to cycles of partial melting and recrystallization. Thus the synchronization process does not completely stop as in the case of the experiment.

\section{Discussion and Conclusion}
The good agreement of the simulation (fig.~\ref{fig:3}) with the experiment (fig.~\ref{fig:2}) suggests that the confinement asymmetry can be used to explain the observed anisotropic triggering of MCI and the synchronization process. The anisotropy of the spectral intensity of the particle velocity fluctuations indicate undoubtedly that the MCI is sensitive to a weak anisotropy in the horizontal confinement. The dispersion relations for a sheared crystal were examined theoretically in refs.~\cite{ivlev2015, zhdanov2015}.

In the experiment, the hot spot at $\theta=60\degree$ is much brighter than the one in the opposite direction $\theta =240\degree$ which almost vanishes, see fig.~\ref{fig:2}(c). This effect --- which is much weaker in the simulations --- may be due to further anisotropies in the crystal structure, stemming for example from defect chains near the boundary of the plasma crystal. It will be subject to further studies.

Confining a plasma crystal in the horizontal plane always makes it internally inhomogeneous. A parabolic confinement is often used in the literature, see, \emph{e.g.}, \cite{totsuji2001, ivlev2003, durniak2011, durniak2013}. The scaling laws of plasma crystals are also well known. In particular, the interaction range $\kappa=a/\lambda$ of such clusters is weakly depended on the strength of horizontal confinement parameter $\Omega_c$,
\begin{equation}\label{eq:kappa}
 \kappa \propto \Omega_c^{-1/2} \,,
\end{equation}
as is easy to verify by using refs.~\cite{peeters1987, totsuji2001, zhdanov2011spontaneous}. Taking into account that the critical vertical confinement for MCI to be triggered, $\Omega_{z, \mathrm{crit}}$, is known to be strongly dependent on the particle interaction range $\kappa$ (see ref.~\cite{couedel2011} for details), the influence of the horizontal confinement strength becomes apparent. For plasma clusters, the dependence of $\Omega_{z, \mathrm{crit}}$ on  $\kappa$ is described by \cite{couedel2011}
\begin{equation}\label{eq:critical}
 \frac{\Omega_{z, \mathrm{crit}}^2}{\Upsilon(\kappa)} \simeq \mathrm{const} \,, 
 \quad \Upsilon(\kappa) = \frac{\kappa^2 + 3\kappa + 3}{\kappa^3} e^{-\kappa} \,.
\end{equation}
The dependence of $\Omega_{z, \mathrm{crit}}$ on $\Omega_c$ can be calculated by combining eqs.~(\ref{eq:kappa}) and (\ref{eq:critical}): 
\begin{equation}\label{eq:FirstVariation}
 \frac{\delta\Omega_{z, \mathrm{crit}}}{\Omega_{z, \mathrm{crit}}} = 
  \Lambda(\kappa) \frac{\delta\Omega_c}{\Omega_c} \,, 
 \quad \Lambda(\kappa) = \frac{1}{4} \left(\kappa + \frac{(\kappa + 3)^2}{\kappa^2 + 3\kappa + 3} \right) .
\end{equation}  
Since $\Lambda(\kappa)\sim 1$ at $\kappa\sim 1$, the relative variation of the instability threshold is practically proportional to the relative variation of the horizontal confinement strength.

The prediction of eccentricity $\epsilon$ due to an anisotropic horizontal confinement can be deduced from eq.~(\ref{eq:kappa}), yielding 
$\epsilon^{\mathrm{theory}}
 = \left[ 1 - (  \kappa(\Omega_\parallel) / \kappa(\Omega_\perp)   )^2   \right]^{1/2}
 \simeq 0.35 \,.$
Of course, in a more detailed analysis the orientation of the anisotropy would have to be taken into account, as the compressibility of the crystal depends on it. Still, this estimate is not far from the values of experiment and simulations. 

The role of frequency synchronization \cite{couedel2014} was not studied here since special care was taken to quantify the orientations of the phase synchronization processes. The interplay of phase and frequency synchronization during the onset of MCI is an important point in the understanding of the collective phenomenon. 

An order parameter motivated by the Kuramoto model is often used to quantify synchronization processes \cite{fukuda2005}. If the interaction is repulsive, complex patterns can arise that call for a detailed analysis. For example, traveling waves \cite{hong2011} or competing domains of different chirality \cite{giver2011} were observed. Here, a local order parameter was proposed which is sensitive to the orientation of the observed synchronization patterns. 

To conclude, it was shown in simulations that an anisotropy of the horizontal confinement can cause an asymmetric triggering of MCI which is accompanied by particle chains with synchronized motion. \revision{To the best of our knowledge, it is reported for the first time that a horizontal compression in simulations of a plasma crystal reproduces well the synchronization process observed in experiments.} For an appropriate orientation of the anisotropy, MCI is triggered in two directions which leads to competing synchronization patterns. If MCI is triggered in one direction, a single pattern dominates. \revision{A new order parameter was proposed that is able to quantify direction-dependent synchronization. We were thus able to identify synchronization patterns that show a pronounced anisotropy.} 

\begin{acknowledgments}
This project received funds from the German Federal Ministry for Economy and Technology under grant number 50WM1441. S.~Zh.\ received support from the European Research Council under the European Union's Seventh Framework Programme (FP7/2007-2013)/ERC Grant agreement 267499. L.~C., \revision{V.~N.\ and S.~Zh.}\ received support from the French-German PHC PROCOPE program (No.\ 28444XH/55926142).
\end{acknowledgments}

\bibliographystyle{eplbib}
\bibliography{literature_sync}

\end{document}